\documentstyle[11pt,newpasp,twoside,epsf]{article}
\def\edcomment#1{\iffalse\marginpar{\raggedright\sl#1\/}\else\relax\fi}
\reversemarginpar

\begin{document}
\title{Kinematic Properties of the TDG Candidates of\\ CG J1720-67.8}
 \author{Sonia Temporin}
\affil{Institute of Astrophysics, University of Innsbruck, Austria}
%\author{Co-authors}
%\affil{Their affiliation}

\begin{abstract}
I use integral field and long-slit spectra from the AAT and 
the Magellan telescope to investigate the
kinematics of several clumps, recently identified 
along the prominent tidal tail of CG J1720-67.8 and suggested as possile 
tidal dwarf galaxy (TDG) candidates.
%Indications of kinematically distinct structures along the tidal tail are found. 
A comparison of photometric and spectroscopic data with evolutionary synthesis 
models suggests burst ages of $\sim$ 6 Myr  for these clumps/TDG candidates.
\end{abstract}
\section{Introduction}

Tidal tails in interacting/merging systems are sites of super-star cluster
and dwarf galaxy formation. The dense environment of compact galaxy groups
offers an extraordinary chance to observe this phenomenon. 
Here I investigate the kinematics of the prominent tidal tail of the
compact galaxy group CG J1720-67.8 through integral field spectra obtained at
the Anglo-Australian Telescope (AAO) and a long-slit spectrum, kindly provided
by F. Kerber, obtained at the Magellan telescope (LCO). This study complements
a previous analysis (Temporin et al. 2003a,b) of the TDG
candidates identified in this group.

\section{Properties of TDG candidates}

\noindent \emph{Kinematics and Dynamics--}
The prominent tidal tail of CG J1720-67.8 shows a general trend of decreasing
radial velocity from north to south with a maximum velocity difference of
$\sim 400$ km s$^{-1}$, as we found from descrete measurements of the clumps
along it (Temporin et al. 2003a). The TDG candidate ``3+9'' at the northern 
tip of the tail (Fig. 1a) shows a steeper velocity gradient (Fig. 1b): 150 to 200 km s$^{-1}$ on
a projected extent of $\sim$ 5 kpc (H$_0$ = 75 km s$^{-1}$ Mpc$^{-1}$).
The long-slit spectrum taken across the southern/central part of the tail shows 
that beside the already observed general trend, there are local velocity gradients
within individual TDG candidates (Fig. 1c). The lack of a continuous trend between adjacent 
clumps suggests that they might be kinematically distinct structures. The observed
local velocity gradients, of order of 20 km s$^{-1}$ kpc$^{-1}$, are comparable 
to those found in other TDGs 
(Mendes de Oliveira et al. 2001).

\medskip

\noindent \emph{Evolutionary History--}
A previous analysis of optical colors indicated a young burst age for the TDG candidates,
in the range 7 to 20 Myr (Temporin et al. 2003b). By comparing the observations
with SB99 models (Leitherer et al. 1999), I found a good agreement of optical/near-infrared
colors and H$\alpha$ equivalent widths with instantaneous burst
models of age 5.5 to 8.5 Myr, once a correction for internal extinction was applied.
Extinction-corrected spectra are in good agreement with model spectra of age $\sim$
6 Myr and total masses of order 10$^7$ M$_{\odot}$ (Fig. 1d). All pieces of evidence suggest 
that these clumps have bursts much younger than their
parent galaxies, for which ages of $\sim$ 40 to 180 Myr were estimated.

\begin{figure}[h]
\vbox{
%\plottwo{temporin.fig1a.eps}{temporin.fig1b.eps}
\plottwo{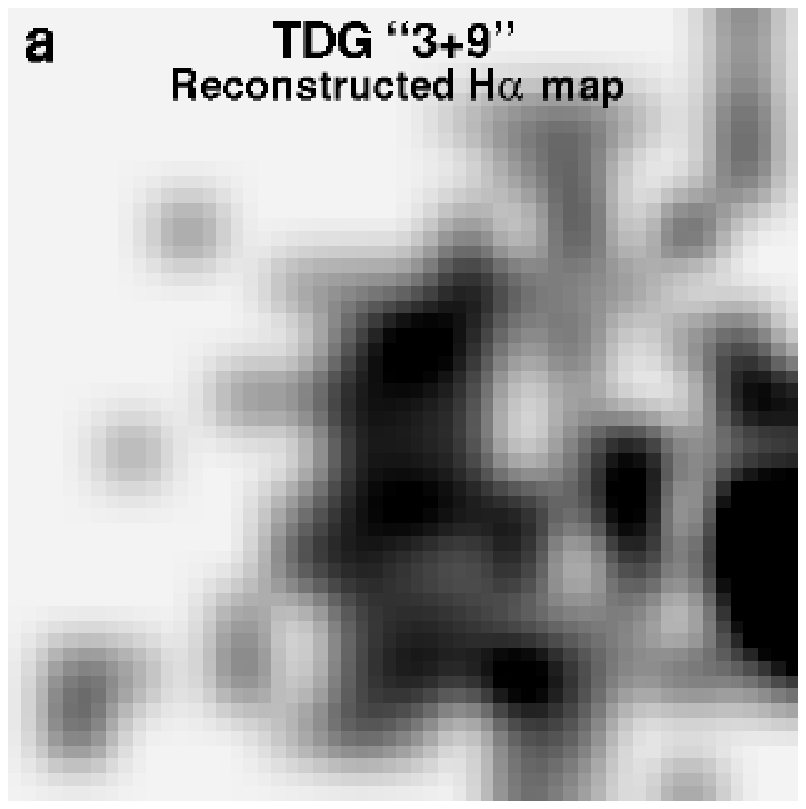}{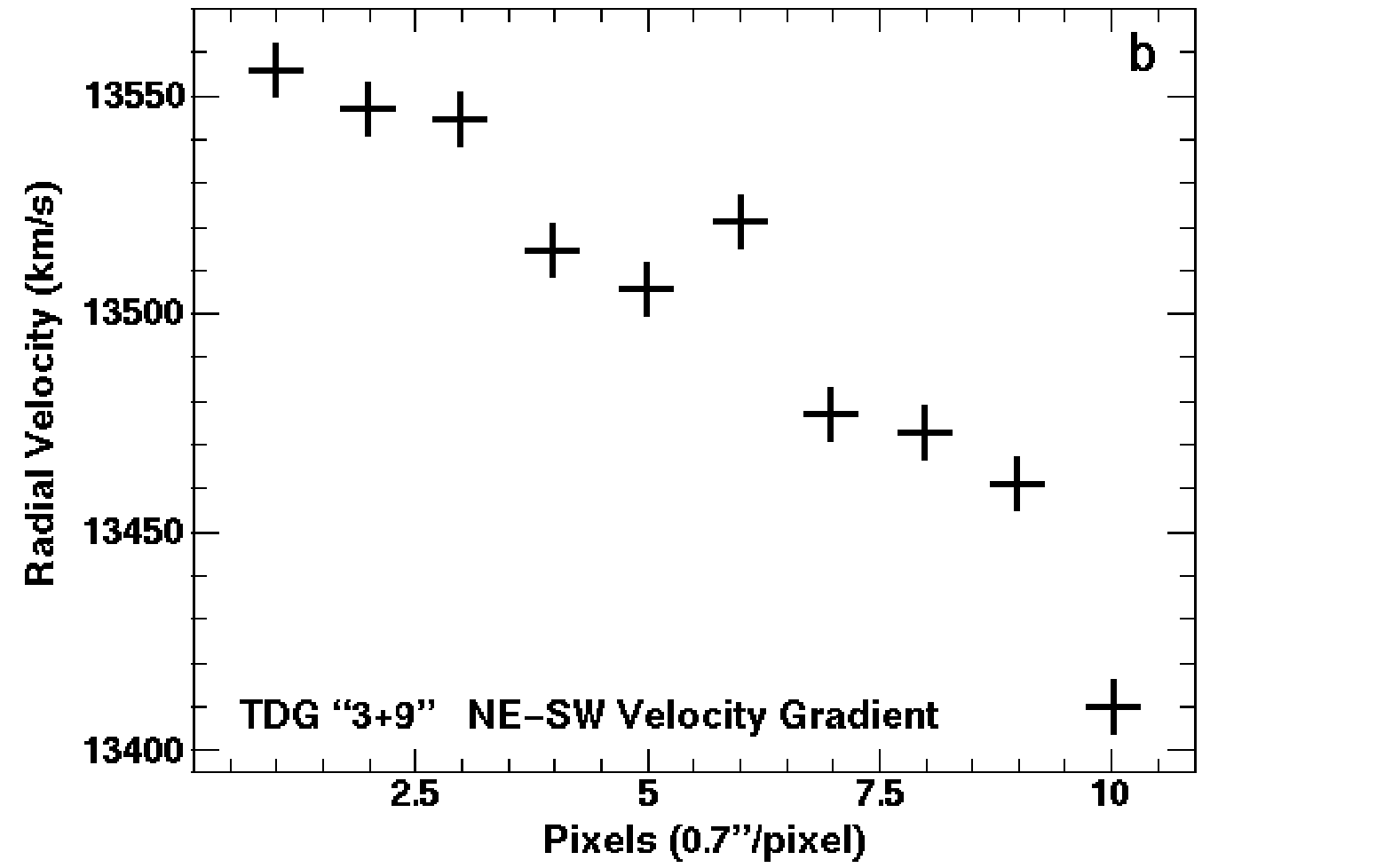}
\plottwo{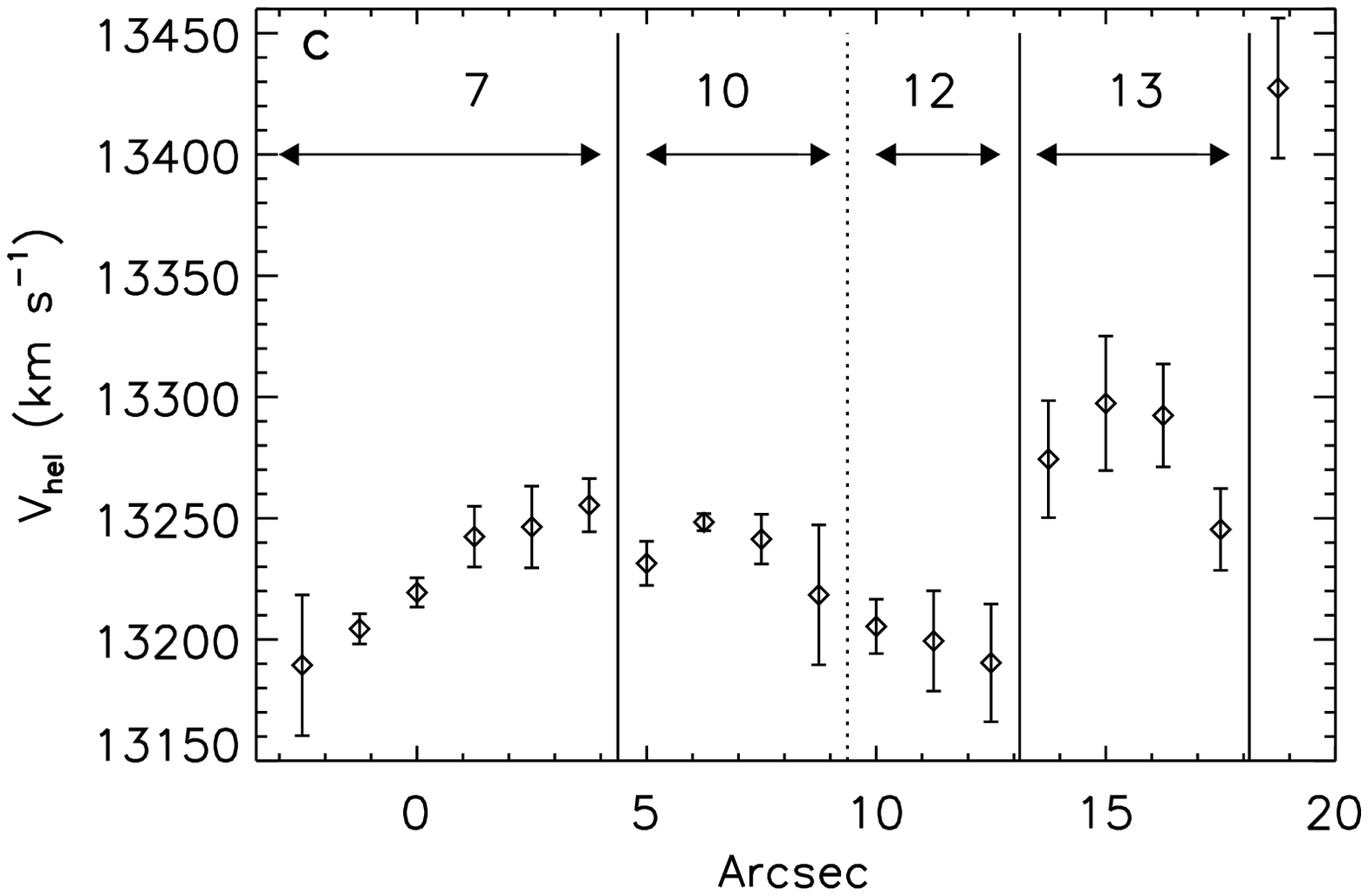}{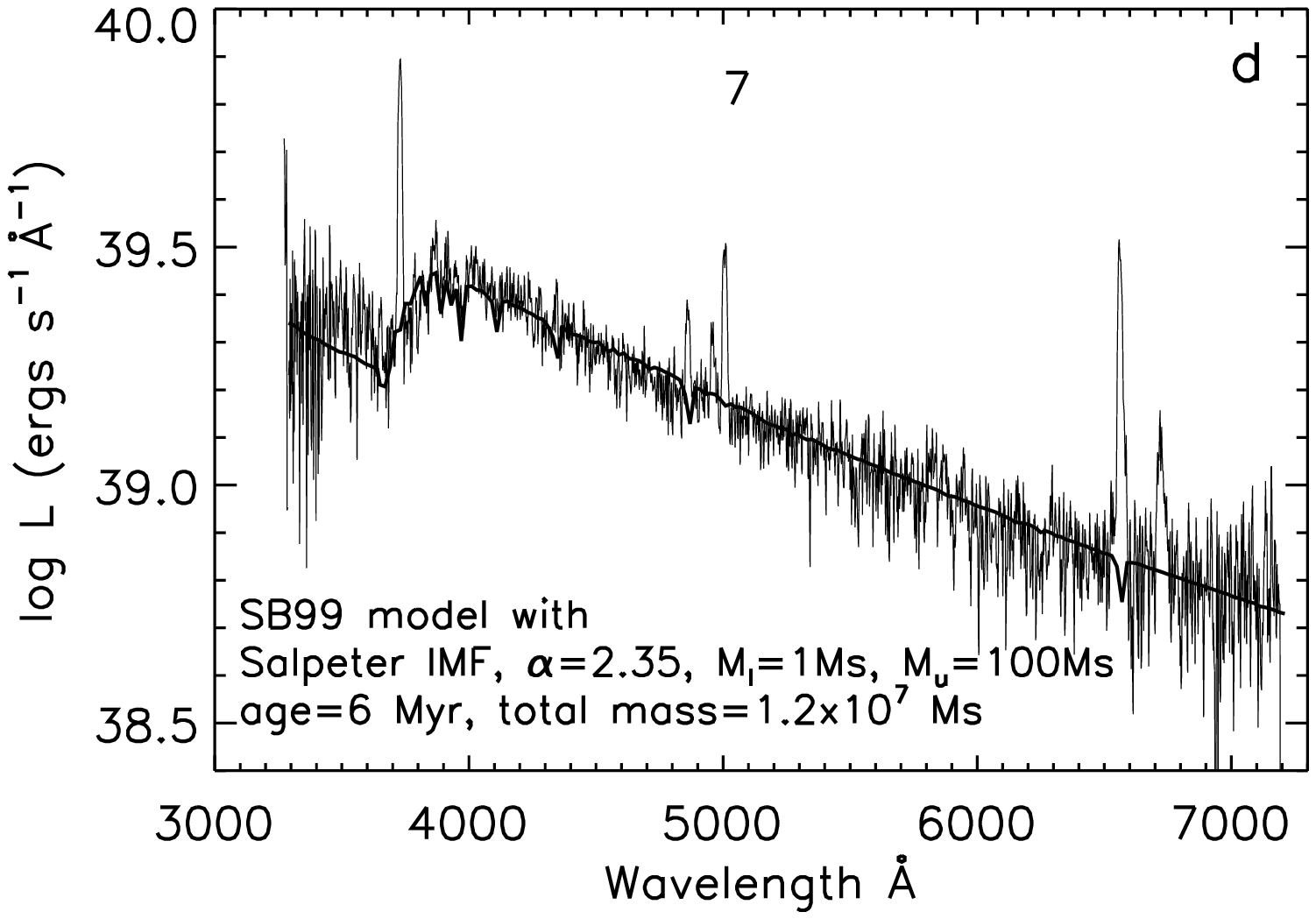}
}
\caption{\emph{a)} H$\alpha$ map of the TDG ``3+9'' reconstructed from integral field data.
\emph{b)} Radial velocity gradient across TDG ``3+9''. \emph{c)} Velocity curve along the
southern part of the tidal tail. Individual clumps are labeled.
\emph{d)} Spectrum of TDG ``7'' compared with an SB99 model, scaled in mass
to match the observed, extinction-corrected luminosity.}
\end{figure}
In \emph{conclusion}, the photometric and spectroscopic properties of these tidal objects
are in agreement with what expected for TDGs. Although projection effects might 
explain part of the observed local velocity gradients, kinematic data seem to 
indicate the presence of self-gravitation in the individual clumps.

\acknowledgments{I am grateful to F. Kerber for providing me with observations
from the Magellan telescope. I acknowledge financial support by the Austrian
Science Fund (FWF) under project P15065.}

\end{document}